# A Reputation Scheme to Discourage Selfish QoS Manipulation in Two-Hop Wireless Relay Networks

Jerzy Konorski
Faculty of Electronics, Telecommunications and Informatics
Gdansk University of Technology
Gdansk, Poland
jekon@eti.pg.edu.pl

Szymon Szott
Faculty of Computer Science, Electronics and Telecommunications
AGH University of Science and Technology
Krakow, Poland
szott@kt.agh.edu.pl

*Abstract*—In wireless networks, stations can improve their received quality of service (QoS) by handling packets of source flows with higher priority. Additionally, in cooperative relay networks, the relays can handle transit flows with lower priority. We use game theory to model a two-hop relay network where each of the two involved stations can commit such selfish QoS manipulation. We design and evaluate a reputation-based incentive scheme called RISC2WIN, whereby a trusted third party (e.g., an access point) can limit selfish behavior and preserve appropriate QoS for both stations.

## I. INTRODUCTION

The distributed nature of wireless local area networks (WLANs) requires for stations to compete for access to a common resource: the radio channel. Stations can improve their level of quality of service (QoS) by handling packets of source flows with higher priority. Additionally, if a station's role in the network involves forwarding traffic, it can handle transit flows with lower priority [1]. We refer to the traffic flow priority upgrading and downgrading behavior as selfish QoS manipulation (SQM) attacks. For a TCP/IP 802.11-based WLAN, examples of mechanisms used to execute an SQM attack would be changing the QoS designation (the Class of Service, CoS) in an IP header or modifying the parameters of the medium access function. Selfish attacks in WLANs have been studied in the literature [1]-[4] and they have been shown to be a threat to two-hop relay networks [5], where cooperation is particularly required.

Two-hop relay networks are a form of cooperative wireless communication used to extend the coverage of WLANs. Stations with a direct connection to an access point (*AP*) act as relays, i.e., share their connection with other, neighboring stations, which either cannot reach the *AP* themselves or have a poor direct connection to it. Such an approach is known to have many advantages in terms of network coverage and performance [6]. Consider the network in Fig. 1, which features a fixed access point (*AP*) and user stations *A* and *B*, with the *AP* and station *A* as well as stations *A* and *B* mutually in range, *AP* and station *B* out of range, and station *A* serving as a voluntary relay for station *B*. For this service station *A* is granted a privileged status at the *AP* (e.g., free or high-speed Internet access). However, without any enforcement, a dominant strategy for a selfish station *A* would be to ensure the lowest possible QoS for station *B* that permits to retain the privileged status (via downgrading SQM attacks such as medium access priority manipulation performed on transit traffic from station *B*), as well as to demand the highest possible QoS for itself (via upgrading SQM attacks such as falsely announcing high CoS performed on own source traffic). Such SQM attacks would ruin station *B*'s QoS perception. On its part, station *B* may find it beneficial to always or frequently demand high QoS by setting high CoS, i.e., by performing upgrading SQM attacks on its source traffic.

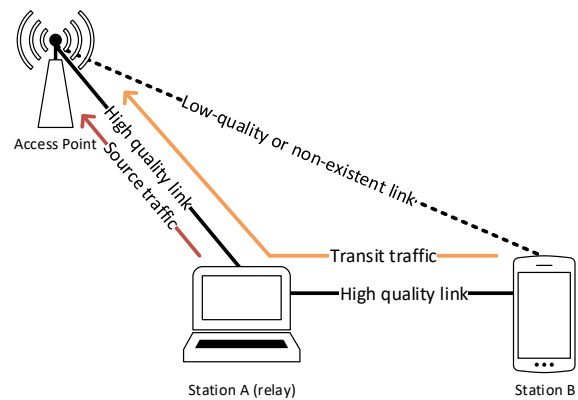

Fig. 1. Cooperative two-hop wireless relay network. From the relay's perspective, uplink traffic is either source or transit.

We assume that deterministic detection of false CoS announcement by examining intrinsic CoS at the *AP*, e.g., using traffic classification [7], is too costly to be practical, and that the intrinsic CoS statistics of generated traffic are not known to other stations, which rules out statistical detection (Section III). Therefore, incentive-based solutions have to be designed. The main challenge is to disincentivize false CoS announcement at stations *A* and *B* (i.e., encourage them to demand QoS in accordance with source traffic's intrinsic CoS) and medium access priority manipulation at station *A* (i.e., encourage it to relay transit traffic from *B* in accordance with the demanded QoS as announced through its CoS). Moreover, station *A* should not find itself forced to ensure a higher QoS for station *B* than it does for itself, in which case it might reconsider trading QoS received by its source traffic for the privileged status at the *AP*. Possible approaches include the following:

- Arrange a scheme whereby station *B* provides some reward to station *A* (monetary or otherwise) depending on the demanded and received QoS. This renders SQM attacks pointless on both sides, but is costly to implement in a realistic scenario (due to the complexity of proper accounting, cryp-

The work of Jerzy Konorski is funded by the National Science Center, Poland, under Grant UMO-2016/21/B/ST6/03146. The work of Szymon Szott is supported by AGH University (contract no. 11.11.230.018).

tographic protocols, fraud prevention, etc.) and may need delicate contract design [8], [9].

- Have the *AP* reward station *A* for ensuring satisfactory, rather than lowest possible, QoS for station *B* in return for the privileged status. However, a fair payment scheme is difficult to design, since QoS demanded by station *B* for own source traffic may result from upgrading SQM attacks.

We take another approach by designing a heuristic reputation scheme at the *AP* that links the prospect of retaining the privileged status by station *A* with receiving satisfactory QoS by station *B*'s source traffic. Our contribution is called *Reputation-based Incentive Scheme for Cooperative 2-hop WIreless relay Networks* (RISC2WIN). In contrast with known reputation schemes designed for distributed wireless networks [10] ours is the first to address cooperative scenarios featuring an *AP* and stations executing SQM attacks (Section III). In RISC2WIN, the *AP* maintains *A*'s reputation $r_A$ such that

- station *A* is interested in keeping $r_A$ away from a critically low value at which its privileged status at the *AP* is revoked,
- acquiring high $r_A$ instills the *AP*'s trust and is beneficial in terms of station *A*'s received QoS, and
- high QoS received by station *B* may bolster $r_A$, while high QoS received by station *A* may lower $r_A$.

Hence the introduced reputation produces desirable effects for station *B* as well as is desired by station *A*; as such it can be compared to commodity, rather than fiduciary, currency.

RISC2WIN operates independently of the underlying QoS provisioning and medium access mechanisms, hence of the wireless technology. It instills a noncooperative game among stations *A* and *B*, where a station's utility is related to received QoS and whose Nash equilibria (NE) yield satisfactory QoS for both stations[1]. A distinctive feature is that the scheme does not rely on SQM attack detection, since the *AP* is either uncertain that an SQM attack is in progress or uncertain that it deserves punishment (Sections III.C and III.D). We define rational threshold-based station strategies (Section IV) and show through simulations that RISC2WIN can limit selfish behavior and preserve QoS for both stations (Section V).

## II. MODEL

### A. QoS Provisioning

Although it can work with any wireless technology, the proposed scheme will be described in the context of IEEE 802.11 networks. QoS differentiation in such networks uses a class-based approach [11]. First, at the IP layer, packets are assigned a CoS which is stored in their IP headers. Then, based on these values the MAC layer, using the enhanced distributed channel access (EDCA) function of IEEE 802.11, maps the higher-layer traffic class to one of the defined access categories (ACs). Without loss of generality, we restrict our analysis to one high-priority AC (voice, *VO*) and one low-priority AC (best effort, *BE*). Each AC is characterized by medium access parameters which assure statistical prioritization with respect to medium access delays. The CoS and AC settings will become important when characterizing a station's traffic and discussing its possible behaviors and strategies.

### B. Traffic

Stations *A* and *B* transmit user *sessions* towards the *AP*. Each session consists of an integer number of *chunks* of fixed duration. We consider a discrete time axis whose each time slot corresponds to a chunk (Fig. 2). Let $\tau_X(k)$, $L_X(k)$, and $iCoS_X(k)$ be, respectively, the start time, duration, and intrinsic class of service of station *X*'s $k$th session, $X \in \{A,B\}$, $k \in \{1,2,\ldots\}$; its successive chunks therefore end at times $\tau_X(k) + 1,\ldots,\tau_X(k) + L_X(k)$. Stations *A* and *B* operate under saturation traffic, i.e., $\tau_X(k) = \tau_X(k-1) + L_X(k)$ (this assumption simplifies the model and can be easily relaxed). Both $L_X(k) \in \{1,2,\ldots\}$ and $iCoS_X(k) \in \{BE,VO\}$ are discrete-time user processes governed by some stationary probability distributions; let $\rho_X = \Pr[iCoS_X(k) = VO]$ ($\rho_A$ and $\rho_B$ are the respective stations' private knowledge). By $CoS_X(k)$ we denote the CoS announced in the IP headers.

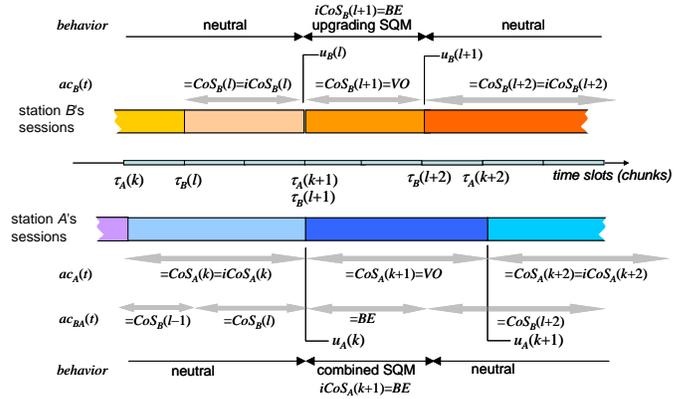

Fig. 2. Traffic model and station behavior.

### C. Behaviors and Strategies

Station *behavior* refers to handling source packets depending on their intrinsic CoS and (in the case of station *A*) handling transit packets depending on their CoS announced in IP headers. At the start of its $k$th session at time $t$, each station *X* demands a certain QoS level by setting $CoS_X(k)$ and decides its medium access priority by setting $ac_X(t)$. Moreover, station *A* decides the medium access priority for transit traffic from station *B* by setting $ac_{BA}(t)$.

*Neutral* behavior prescribes setting $CoS_X(k) = iCoS_X(k)$, $ac_X(t) = CoS_X(k)$, and (at station *A*) $ac_{BA}(t) = CoS_B(k)$. However, announcing a different CoS (i.e., false CoS announcement) and deciding a different medium access priority is possible and subject to station *X*'s discretion. We assume that station *X* cannot falsify session delimiters, i.e., pretend that the $k$th session starts at a time different from $\tau_X(k)$. Session delimiters are also assumed to be recognizable to both stations and the *AP*. In particular this implies that $CoS_X(k)$ remains unchanged throughout the $k$th session. On the other hand, $ac_X(t)$ and $ac_{BA}(t)$ can be decided on a chunk-by-chunk basis; to avoid excessive decision making we assume that they can only be changed at any station's session ends. Any behavior other than neutral will be referred to as an *SQM attack*.

---
[1] An NE is an operating point at which no station can benefit by unilaterally changing its behavior [12].

A station *strategy* refers to a well-planned sequence of behaviors across successive sessions that aims to maximize long-term perception of received QoS, quantified in Section II.E.

### D. Attack Feasibility and Detectability

Incentive-based solutions are only justified when SQM attacks are undetectable outside the attacker station or cannot be fairly punished as discussed later. Since we have assumed that neither the *AP* nor the stations *A* and *B* have a traffic classification capability and that $iCoS_X(k)$ and $\rho_X$ are private knowledge, falsifying CoS headers of source packets is not easily detectable. However, station *A* must refrain from detectable misbehavior toward traffic from station *B* as it would jeopardize its privileged status at the *AP*; e.g., packet dropping or falsifying CoS headers could be directly observed at station *B* by sensing station *A*'s transmissions or by communication with the *AP* via a secure end-to-end signaling scheme.

The detectability of SQM attacks through selfish manipulation of medium access priority depends on the attack mechanism used. In the case of EDCA, forcing a packet into a desired AC queue must be accompanied by setting CoS accordingly in the packet header (i.e., $ac_X(t) = CoS_X(k)$ must hold when station *X*'s *k*th session is starting), otherwise the discrepancy between the CoS and AC fields in the packet header is easily observable.[2] Therefore the attack is undetectable if performed on source traffic, but detectable if performed on transit traffic. An undetectable SQM attack is also possible by artificially delaying packets before passing them into an AC queue; note that only downgrading is practical in this case, since upgrading would require that artificial delay be applied by default to all *BE* traffic and only skipped during an attack.

To summarize, the following SQM attacks are both feasible and undetectable:

- falsify CoS and set *ac* accordingly to perform *upgrading* or *downgrading* of source traffic, i.e., $iCoS_X(k) \ne ac_X(t) = CoS_X(k)$ at any station $X \in \{A,B\}$, and

- introduce artificial delay at station *A* to perform downgrading of transit traffic, i.e., $VO = CoS_B(k) \ne ac_{BA}(t) = BE$; if performed along with upgrading of source traffic via CoS falsification, such an SQM attack is referred to as *combined*.

Fig. 2 illustrates station behaviors. Note that although at time $t = \tau_B(l + 2)$ station *A* chooses neutral behavior, $CoS_A(k + 1) \ne iCoS_A(k + 1)$ must be maintained until $t = \tau_A(k + 2)$.

### E. QoS and Utilities

A chunk is assumed long enough for the current behaviors to take effect and for both stations to observe the received QoS with satisfactory accuracy. To this end, station *A* conducts routine measurements of throughput, packet delays and loss, whereas station *B* derives those either by observing station *A*'s transmissions in a single-channel network, or via secure end-to-end signaling from the *AP* (e.g., using a direct low-quality link as in Fig. 1, if it exists, or encrypted transport-layer messages). We moreover assume that received QoS only depends on the ACs of the uplink traffic from stations *A* and *B* competing for bandwidth around station *A* (that is, at the considered level of QoS granularity, we neglect the impact of downlink traffic from the *AP* or hidden stations, if any, as well as of transmission impairments, virtual collisions among AC queues, transport-layer protocols, etc.). The QoS received by station *X*'s source session chunk starting at time *t* can therefore be expressed as $f_X(ac_A(t), ac_{BA}(t), ac_B(t))$, where $f_X(\cdot)$ is a public knowledge function whose exact form depends on the MAC performance model. Note that received QoS is observable to both stations and the *AP*. In our simplified model, only two QoS levels, low and high, are distinguished. The low level, assigned here the numerical value 0, is satisfactory for *BE* traffic (as this traffic class has no QoS requirements, no level below that is defined), and the high level, with the numerical value 1, is satisfactory for *VO* traffic. Hence, $f_X(\cdot) \in \{0, 1\}$. To get a qualitative insight, for $X = A$ we take $f_X(ac_A(t), ac_{BA}(t), ac_B(t)) = 1$ iff $ac_A(t) = VO \land (ac_B(t) = BE \lor ac_{BA}(t) = BE)$, and for $X = B$, $f_X(ac_A(t), ac_{BA}(t), ac_B(t)) = 1$ iff $ac_A(t) = BE \land ac_B(t) = VO \land ac_{BA}(t) = VO$. That is, station *A* receives high-level QoS if its source *VO* traffic does not have to compete with the two segments (incoming and relayed) of the *VO* traffic flow from station *B*, whereas station *B* receives high-level QoS if its source *VO* traffic is relayed as such and does not have to compete with *VO* source traffic from station *A*. With so defined $f_X(\cdot)$, the strategic situation is not quite checks and balances: station *A* has a strategic advantage, as it can always ensure high-level QoS for its source session by performing a combined SQM attack, while station *B* only receives high-level QoS for its source session if station *A* does not demand high-level QoS. Per-session *utility* of station $X \in \{A,B\}$, calculated at the end of the *k*th session, is taken to be the QoS level averaged over all chunks throughout the session duration, cf. Fig. 1:

$$u_X(k) = \frac{\sum_{\tau_X(k-1)<t\le\tau_X(k)} f_X\left(ac_A(t),ac_{BA}(t),ac_B(t)\right)}{\tau_X(k) - \tau_X(k-1)}. \quad (1)$$

Note that since received QoS and session delimiters are observable to both stations and the *AP*, so are $u_X(k)$. For specific sequences $(L_X(k))_{k=1,2,\ldots} = (\tau_X(k) - \tau_X(k-1))_{k=1,2,\ldots}$ and $iCoS_A(k))_{k=1,2,\ldots}$, long-term average utilities of station *X* fall between 0 and 1 and are given by

$$u_X^{ac} = \lim_{K\to\infty} \frac{\sum_{k=1..K: iCoS_X(k)=ac} u_X(k)}{\sum_{k=1..K: iCoS_X(k)=ac} 1}, \quad (2)$$

where $ac \in \{BE,VO\}$. In the case when no SQM attacks are performed by either station, $u_X(k) = 0$ if $iCoS_X(k) = BE$, hence $u_X^{BE}\big|_{\text{noSQM}} = 0$; moreover, $u_X(k) \le 1$ if $iCoS_X(k) = VO$ due to the "natural" competition between overlapping chunks of intrinsically *VO* sessions transmitted by stations *A* and *B*, hence typi-cally $u_X^{VO}|_{noSQM} < 1$. For the saturated traffic scenario and in the absence of incentive schemes, $u_X^{VO}|_{noSQM, noincentives} = u_X^o = 1 - \rho_{X'}$, where $X'$ denotes the other station. If SQM attacks are performed, their success can be measured by $u_X^{BE} > 0$.

---

[2] Undetectable setting of $ac_X(t) \ne CoS_X(k)$ under EDCA would be feasible if AC fields in transmitted packets could be set arbitrarily, i.e., if the MAC firmware could be easily tampered with; another option would be packet-by-packet changing of EDCA parameters of an AC queue. Both options are impractical from an implementation viewpoint.

Each station $X$ plans its strategy with a goal to maximize $U_X = u_X^{BE} + w u_X^{VO}$, where $w > 0$ reflects the importance the station attaches to the QoS received by its source $VO$ traffic relative to the success of SQM attacks applied to its source $BE$ traffic (in a realistic model, $w \geq 1$). Maximizing $U_X$ is easy if the other station exhibits neutral behavior—it is enough to constantly perform an SQM attack: combined if $X = A$, or upgrading of source traffic if $X = B$. Indeed, $f_A(1, 0, ac_B(t)) \geq f_A(ac_A(t), ac_{BA}(t), ac_B(t))$ and $f_B(ac_A(t), 1, 1) \geq f_B(ac_A(t), ac_{BA}(t), ac_B(t))$ regardless of $ac_A(t)$, $ac_{BA}(t)$, and $ac_B(t)$. The other station is then left with zero utilities, since $f_B(1, 0, ac_B(t)) = f_A(ac_A(t), 1, 1) = 0$ regardless of $ac_A(t)$ and $ac_B(t)$. Hence, neutral behavior does not ensure satisfactory utilities in the face of SQM attacks and SQM attacks may have to be performed in self-defense. This sheds new light on SQM attack detectability and punishability. The *AP* might detect and punish an ongoing SQM attack based on observed $u_X(k)$ and the public function $f_X(\cdot)$. For example, $u_B(k) < 1$ or $u_A(l) = 1$ while $CoS_B(k) = VO$ indicates downgrading of transit traffic (and possibly upgrading of source traffic) at station *A*. However, the punishment is not guaranteed to be fair: station *B* may be performing an undetectable upgrading SQM attack and station *A*'s attack may be performed in self-defense. The above difficulty of the detection and punishment approach further supports an incentive-based approach, where stations *A* and *B* are players in a one-shot noncooperative game (whereas the *AP* is a trusted third party).

In the considered game, the stations' long-term average utilities are the payoffs. Due to station *A*'s strategic advantage, without any external incentive scheme the game admits a unique weak NE with highly asymmetric utilities, incentivizing station *A* to select a combined SQM attack for each session and leaving station *B* with a zero utility. This is because $f_A(1, 0, ac_B(t)) = 1$ and $f_B(1, 0, ac_B(t)) = 0$ regardless of $ac_B(t)$.

## III. REPUTATION SCHEME

The proposed RISC2WIN reputation scheme aims to encourage stations' strategies that avoid the above described undesirable Nash equilibrium. Note that it is probably impossible to disincentivize SQM attacks completely if their direct detection and punishment is not employed and if their success is valued highly by the two stations (i.e., if $w$ is small). In line with the previous discussion we postulate that if $w$ is large enough then at the Nash equilibria,

- $u_X^{BE}$ is relatively small, i.e., station $X$ attacks with restraint,

- $u_X^{VO}$ is not distinctly less than $u_X^o$, i.e., QoS received for intrinsically $VO$ traffic should only depend on the "natural" competition between $VO$ traffic from stations *A* and *B*, and

- station *A* typically does not receive lower QoS for its source traffic than does station *B*, i.e., $u_A^{ac} \geq u_B^{ac}$ for $ac \in \{BE, VO\}$.

To reward station *A* for its relay services, the *AP* maintains station *A*'s current reputation $r_A$, and updates it session by session based on both stations' received QoS (more precisely, per-session utility). The idea is for station *A* to keep $r_A$ within an acceptable range. We prescribe that $r_A$ is incremented when station *B*'s *VO* traffic receives, or station *A*'s *VO* traffic does not receive 'high enough' QoS, and decremented when station *B*'s *VO* traffic does not receive, or station *A*'s *VO* traffic does receive 'high enough' QoS. The QoS level that counts as 'high enough' is the lower, the higher the current $r_A$. Accordingly, the reputation scheme follows principles (i) through (v):

(i) to retain a privileged status at the *AP*, station *A* must not let $r_A$ drop below a critical value, here fixed at 0; for convenience, we allow only discrete $r_A \in \{0, 1, \dots, R\}$ with $R \geq 1$,

(ii) to acquire a higher $r_A$, station *A* must ensure that station *B* frequently receives high-level QoS if it demands it for its source traffic,

(iii) station *A* receiving high QoS for its source traffic retains $r_A$ if station *B* is demanding high QoS,

(iv) at a higher $r_A$, station *A* finds it easier to receive high-level QoS for its source traffic, while caring less for the QoS received by station *B*, and

(v) by demanding high QoS with restraint station *B* curbs $r_A$.

Thus a high $r_A$ translates for station *A* into the *AP*'s trust that manifests itself in more liberal conditions of keeping $r_A$ high in the future. This in turn improves station *A*'s QoS-related utility. One notices that the above design principles do not explicitly refer to SQM attack detection.

Let $r_A(t)$ be station *A*'s current reputation at time $t = 0, 1, 2, \dots$ and $\bar{u}_X(t)$ be the average station *X*'s per-session utility (including *BE* and *VO* traffic) observed by the *AP* up to time $t$. Define the modified reputation as

$$r_{Am}(t) = \begin{cases} \min\left\{1, (r_A(t)+1)\frac{\bar{u}_B(t)}{\bar{u}_A(t)} - 1\right\}, & \bar{u}_B(t) > \bar{u}_A(t) \\ r_A(t), & \text{otherwise.} \end{cases} \quad (3)$$

Any changes in $r_A$ can be made only at the times when station *A*'s or station *B*'s session ends. Consider first $t = \tau_B(k+1)$, where $\tau_A(l) < \tau_B(k+1) \leq \tau_A(l+1)$, i.e., station *B*'s $k$th session has just ended and station *A*'s $l$th session is in progress, and denote by $t^-$ the instant of time just prior to $t$. The *AP* can read the demanded QoS level as $CoS_B(k)$ in packet headers transmitted by station *B* and faithfully relayed by station *A* as argued in Section II.D (recall that $CoS_B(k)$ is subject to station *B*'s behavior choice and need not equal $iCoS_B(k)$). If $CoS_B(k) = BE$, $r_A(t^-)$ is not incremented (as dictated by principle (v)) or decremented (for otherwise station *B* might too easily damage $r_A(t)$ by performing downgrading SQM attacks). In line with principle (ii), if $CoS_B(k) = VO$ then the *AP* checks if the QoS received by station *B* is 'high enough':

$$u_B(k) \geq 1 - r_{Am}(t^-)/R, \quad (4)$$

in which case $r_A(t^-)$ is incremented by 1. Otherwise $r_A(t^-)$ is retained rather than decremented—such a design decision stems from the possibility that $CoS_B(k) = VO$ signifies an ongoing upgrading SQM attack at station $B$ or that station $A$ may be transmitting legitimate $VO$ source traffic ($CoS_A(l) = iCoS_A(l)$) and not performing an upgrading SQM attack. In line with principle (iv), the closer $r_A(t)$ is to $R$, the easier it is for station $A$ to produce a satisfactory comparison at the $AP$ and drive its reputation away from the critical value 0. This permits to receive high QoS for future source traffic (perhaps through SQM attacks) with less concern about producing satisfactory comparisons. Due to (3), a satisfactory comparison is also easier to produce when $\bar{u}_B(t) > \bar{u}_A(t)$; this heuristic adds an element of self-regulation: station $A$ regains high reputation faster if recent reputation dynamics have been favoring station $B$'s utilities.

Consider now $t = \tau_A(k + 1)$, where $\tau_B(l) < \tau_A(k + 1) \leq \tau_B(l + 1)$, i.e., station $A$'s $k$th session has just ended and station $B$'s $l$th session is in progress. Recall that station $A$ signals the demanded QoS level as $CoS_A(k)$ in transmitted packet headers ($CoS_A(k)$ is subject to a behavior choice and need not equal $iCoS_A(k)$). If $CoS_A(k) = BE$, no changes of the current reputation are made. If $CoS_A(k) = VO$ then the $AP$ checks if

$$u_A(k) \geq r_{Am}(t^-)/R, \quad (5)$$

in which case $r_A(t^-)$ is either decremented by 1 if $CoS_B(l) = BE$, or retained if $CoS_B(l) = VO$. Thus receiving relatively low QoS cannot damage station $A$'s reputation and, in line with principle (iii), neither can receiving high QoS if at the same time station $B$ is demanding high QoS. On the other hand, in line with principle (v), the reputation may go down if station $A$ receives high QoS while station $B$ is demanding low QoS. Such design decisions are to disincentivize upgrading SQM attacks at station $B$. As before, the self-regulation heuristic (3) also applies. Thus the reputation dynamics are: at $t = \tau_B(k + 1)$, $r_A(t)$ becomes $\min\{R, r_A(t^-) + 1\}$ iff

$$CoS_B(k) = VO \wedge u_B(k) \geq 1 - r_{Am}(t^-)/R, \quad (6)$$

whereas at $t = \tau_A(k + 1)$ such that $\tau_B(l) < \tau_A(k + 1) \leq \tau_B(l + 1)$, $r_A(t)$ becomes $r_A(t^-) - 1$ iff

$$CoS_A(k) = VO \wedge u_A(k) \geq r_{Am}(t^-)/R \wedge CoS_B(l) = BE, \quad (7)$$

and in all other situations remains unchanged. If $\tau_A(k + 1) = \tau_B(l + 1) = t$, i.e., both stations' session ends coincide on the slotted time axis, then (6) is checked first and the resulting $r_A(t)$ is substituted for $r_A(t^-)$ when checking (7).

The updated reputation values are disseminated by the $AP$ using a direct low-quality link (Fig. 1), if it exists; even if it does not, then, provided that $r_A(0)$ is public knowledge, stations $A$ and $B$ can follow the above dynamics as they can both observe $CoS_X(k)$ and $u_X(k)$. Hence $r_A(k)$, too, is public knowledge. This gives rise to the heuristic strategies described below.

IV. THRESHOLD STRATEGIES

A truly sophisticated station $X$'s strategy would determine its behavior choice for each new session based on the whole or recent history of both stations' observed behaviors and utilities, with a view of maximizing the long-term average utility $U_X$. This might be computationally quite complex and require far-sighted discounting of future per-session utilities. For relatively inexpensive wireless devices, simpler strategies should be envisaged that do not slow down packet transmission and only require a few arithmetic operations per session. In this study we narrow the class of considered stations strategies down to memoryless reputation-driven (that is, the history of observed behaviors and utilities is only accounted for through the current reputation value). As discussed in Section II.C, $CoS_X(k)$ is fixed throughout station $X$'s $k$th session, and $ac_X(t)$ and $ac_{BA}(t)$ can only change at either station's session ends. New behaviors are set at either station's session ends, following (6) or (7).

Station $B$'s strategy is represented as $s_B$: $\{BE, VO\} \times \{0,1,\ldots,R\} \rightarrow \{BE, VO\}^2$; depending on $iCoS_B(k)$ and $r_A(t^-)$ it determines $CoS_B(k)$ at times $t = \tau_B(k)$, and $ac_B(t)$ at times $t = \tau_B(k)$ or $t = \tau_A(l)$. Given the proposed reputation scheme, a reasonable heuristic strategy prescribes an upgrading SQM attack if $r_A(t^-)$ is below a predefined threshold $T_{B,up}$, so that there is no imminent threat of station $A$ acquiring a reputation level that would lessen its concern for station $B$'s QoS. This means that station $B$ temporarily gives precedence to receiving high QoS over curbing station $A$'s reputation. On the other hand, if $r_A(t^-)$ is above a predefined threshold $T_{B,down}$, station $B$ attempts to curb station $A$'s rising reputation by performing a downgrading SQM attack on own source traffic. In this way, station $A$ is not given the opportunity to raise its reputation according to (6). In other cases, station $B$ behaves neutrally. Thus $s_B$ is fully characterized by the pair ($T_{B,down}$, $T_{B,up}$), where $T_{B,down} \geq T_{B,up}$.

Station $A$'s strategy is represented as $s_A$: $\{BE, VO\} \times \{0,1,\ldots,R\} \rightarrow \{BE, VO\}^3$; depending on $iCoS_A(k)$ and $r_A(t^-)$ it determines $CoS_A(k)$ at times $t = \tau_A(k)$, and $ac_A(t)$ and $ac_{BA}(t)$ at times $t = \tau_A(k)$ or $t = \tau_B(l)$. Analogously to $s_B$, it prescribes a combined SQM attack (an upgrading SQM attack performed on own source traffic and a downgrading SQM attack performed on the transit traffic from station $B$) if $r_A(t^-)$ is above a threshold $T_{A,comb}$, a downgrading SQM attack performed on own source traffic if $r_A(t^-)$ is below a threshold $T_{A,down}$, and neutral behavior otherwise. This means that when $r_A(t^-)$ is high, station $A$ temporarily gives precedence to receiving high QoS over keeping the reputation high, and when $r_A(t^-)$ is low, keeping it away from 0 is a priority. Thus $s_A$ is fully characterized by the pair ($T_{A,comb}$, $T_{A,down}$), where $T_{A,comb} \geq T_{A,down}$.

Let $t^+$ denote the instant of time immediately after $t$. Stations' strategies are formally specified as follows: at $t = \tau_B(k)$ such that $\tau_A(l - 1) < \tau_B(k) \leq \tau_A(l)$,

$$(ac_A(t^+), ac_{BA}(t^+))$$
$$= \begin{cases} (BE, CoS_B(k)), & r_A(t) \leq T_{A,down} \\ (CoS_A(l), BE), & r_A(t) \geq T_{A,comb} \\ (CoS_A(l), CoS_B(k)), & \text{otherwise}, \end{cases} \quad (8)$$

$$(CoS_B(k), ac_B(t^+))$$
$$= \begin{cases} (BE, BE), & r_A(t) \geq T_{B,down} \\ (VO, VO), & r_A(t) \leq T_{B,up} \\ (iCoS_B(k), iCoS_B(k)), & \text{otherwise}, \end{cases} \quad (9)$$

and at $t = \tau_A(k)$ such that $\tau_B(l - 1) < \tau_A(k) \leq \tau_B(l)$,

$$ac_B(t^+) = \begin{cases} BE, & r_A(t) \geq T_{B,down} \\ CoS_B(l), & \text{otherwise,} \end{cases} \quad (10)$$

$$(CoS_A(k), ac_A(t^+), ac_{BA}(t^+))$$
$$= \begin{cases} (BE, BE, CoS_B(l)), & r_A(t) \leq T_{A,down} \\ (VO, VO, BE), & r_A(t) \geq T_{A,comb} \\ (iCoS_A(k), iCoS_A(k), CoS_B(l)), & \text{otherwise.} \end{cases} \quad (11)$$

By convention, if $T_{A,comb} = T_{A,down}$ or $T_{B,up} = T_{B,down}$, the upper parts of (8)-(11) are checked first, i.e., when arbitrating between a downgrading and an upgrading or combined SQM attack, the former is favored. In the upper part of (9), station $B$ refrains from an upgrading SQM attack (i.e., setting $CoS_B(k) = VO$) when $r_A$ is high to prevent its further incrementing according to (6); as discussed in Section III.C, setting $ac_B(t) = VO$ is then impossible. In the middle part of (9), a downgrading SQM attack (i.e., setting $CoS_B(k) = BE$) is pointless when $r_A$ is low, since it would imply $ac_B(t) = BE$ and unnecessary loss of QoS. In the middle part of (11), station $A$ does not refrain from a combined SQM attack when $r_A$ is high for similar reasons. At a low $r_A$, station $A$ cannot set $ac_A(t) = VO$ without exposing $r_A$ to a decrement, since $CoS_A(k) = VO$ would also have to be set. Likewise, station $B$ cannot receive high QoS while exposing station $A$ to a reputation decrement, since $ac_B(t) = VO$ would imply $CoS_B(k) = VO$ and (7) would not hold.

When seeking high utilities $U_A$ and $U_B$, stations $A$ and $B$ engage in a one-shot noncooperative game $\langle \{A,B\}, S_A \times S_B, (U_A, U_B) \rangle$, where the set $S_X$ of feasible station $X$'s strategies consists of threshold pairs. $(T_{A,comb}, T_{A,down}, T_{B,up}, T_{B,down})$ represents a strategy profile. Clearly, $T_{A,down} = R$ or $T_{B,down} = 0$ are uninteresting as they imply persistent downgrading SQM attacks performed on source $VO$ traffic, producing $u_X^{VO} = 0$. For any set $Z \subseteq \mathbf{R}$ denote $Z_+ = \{(x,y) \in Z \times Z \mid x \geq y\}$; then $S_A = \{0,1,\ldots,R\}_+ \setminus \{(R,R)\}$ and $S_B = \{0,1,\ldots,R\}_+ \setminus \{(0,0)\}$.

V. PERFORMANCE

To evaluate the effects of the proposed reputation scheme, we have simulated scenarios of the network operation according to our model. User session attributes $(L_X(k))_{k=1,2,\ldots} = (\tau_X(k) - \tau_X(k-1))_{k=1,2,\ldots}$ and $(CoS_X(k))_{k=1,2,\ldots}$ were modeled as iid sequences with $\Pr[L_X(k) = k] \equiv 0.1$ for $6 \leq k \leq 15$, and various $\rho_A$ and $\rho_B$. Per-session utilities $u_X(k)$ for successive sessions as well as long-term average utilities were recorded; the weighted-sum utilities $U_X$ were calculated assuming $w = 10$. Reputation values $r_A(\tau_B(k))$ were also recorded, with $R = 10$ and $r_A(0) = R$ fixed throughout all simulations.

A baseline scenario is when both stations behave neutrally, i.e., $(T_{A,comb}, T_{A,down}, T_{B,down}, T_{B,up}) = (\infty, -\infty, \infty, -\infty)$. One then observes $u_X(k) \equiv 0$ when $CoS_X(k) = BE$ and often $u_X(k) < 1$ when $CoS_X(k) = VO$, the latter on account of "natural" competition of $VO$ sessions from stations $A$ and $B$. Since the received QoS reflects the stochastic sequence of successive sessions of $BE$ and $VO$ traffic, $r_A$ follows a random walk with a reflecting barrier at $r_A = R$; consequently, it typically drops below 0 relatively soon (Fig. 3, *left*). To retain its privileged status at the $AP$, station $A$ applies $T_{A,down} = 0$, producing another reflecting barrier at $r_A = 0$. This typically has little bearing upon $u_X^{VO}$ (Fig. 3, *middle*). Selecting nontrivial thresholds in $S_A \times S_B$ can bring about successful upgrading SQM attacks as well as improve some or both stations' utilities (Fig. 3, *right*). Simulations demonstrate that the utility trajectories are quite sensitive to the relative position of the stations' thresholds; usually $r_A$ is attracted to the interval between neighboring thresholds.

In search of high weighted-sum utilities $U_X$, the two selfish stations can set in motion iterative processes such as best-reply dynamics or reinforcement learning. Under a wide class of game models, such processes are likely to reach a NE of the one-shot noncooperative game [13]. Therefore, even disregarding the specifics of the dynamic play, one can get insight into its likely outcomes by studying the set of NE of the one-shot game. Since the iterations are possibly driven by the successive per-session utilities, which may be observed inaccurately, either station may have difficulty finding an exact best reply to the other station's play. A suitable solution concept is $\varepsilon$-*Nash equilibrium* ($\varepsilon$-NE). In the considered game, a strategy profile $(\hat{s}_A, \hat{s}_B)$ is an $\varepsilon$-NE if either player's utility is within $\varepsilon$ of that corresponding to its best-reply strategy. That is,

$$U_X(\hat{s}_X, \hat{s}_{-X}) \geq (1-\varepsilon) \max_{s_X \in S_X} U_X(s_X, \hat{s}_{-X}) \quad \forall X \in \{A,B\}, \quad (12)$$

where $\varepsilon$ is a small number in [0, 1]. For each feasible strategy profile in $S_A \times S_B$, 10 simulation runs have been conducted with fixed $(\rho_A, \rho_B)$ and using a fixed set of pseudorandom sequences $(L_X(k))_{k=1,2,\ldots}$ and $(CoS_X(k))_{k=1,2,\ldots}$. After a simulation run, the obtained asymptotic utilities were transformed into $U_X = u_X^{BE} + w u_X^{VO}$ and $\varepsilon$-NE were found according to (12). Due to the stochastic nature of the iid sequences, each simulation run produced a different set of $\varepsilon$-NE, albeit with a similar utility range. Fig. 4 depicts the corresponding equilibrium $(u_X^{BE}, u_X^{VO})$ pairs for various $(\rho_A, \rho_B)$. It is visible that:

- the sets of long-term average utility pairs at $\varepsilon$-NE are qualitatively not very sensitive to $(\rho_A, \rho_B)$,

- the concept of $\varepsilon$-NE filters out strategy profiles producing $u_X^{BE} = u_X^{VO} = 0$, i.e., selfish dynamic play is not likely to starve station $A$ or $B$; this is in contrast with the play with no reputation scheme or under ill-chosen strategy profiles,

- when distinctly more importance is attached to $VO$ traffic, i.e., $w > 1$, two types of $\varepsilon$-NE occur: at the more desirable type, $u_X^{BE}$ is close to 0, hence few upgrading SQM attacks are successful, and $u_X^{VO}$ is typically not lower than in the absence of SQM attacks; at the other type, near the diagonal $u_X^{BE} = u_X^{VO}$, the utilities are more balanced, with $u_X^{BE} > 0$ signifying frequent successful SQM attacks, and $u_X^{VO}$ somewhat lower than in the absence of SQM attacks, and

- at $\varepsilon$-NE, $u_A^{ac} > u_B^{ac}$ occurs more often than not; thus, as expected, dynamic rational play permits station $A$ to capitalize

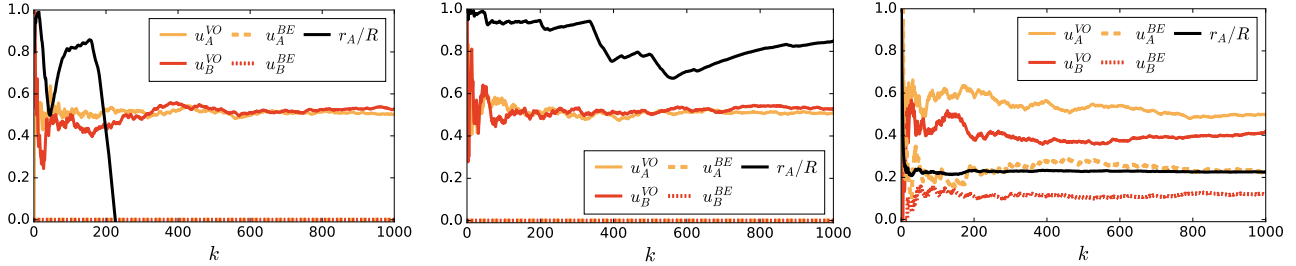

Fig. 3. Utility and reputation trajectories for $(\rho_A, \rho_B) = (0.5, 0.5)$ (with $(u_A^o, u_B^o) = (0.5, 0.5)$) and various strategy profiles $(T_{A,comb}, T_{A,down}, T_{B,down}, T_{B,up})$; *left*: $(\infty, -\infty, \infty, -\infty)$, *middle*: $(\infty, 0, \infty, -\infty)$, *right*: $(3,1,4,1)$; depicted are moving averages with learning constant decaying as $k^{-0.05}$, asymptotic values approximate $u_X^{ac}$. Note that in the left and middle figures, $u_X^{BE}$ assumes only zero values for $X \in \{A, B\}$.

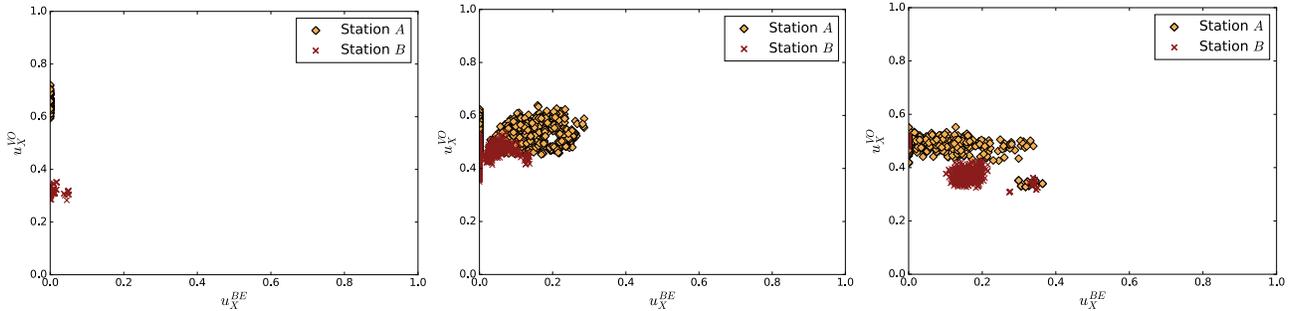

Fig. 4. Stations' long-term average utilities at $\varepsilon$-NE for $\varepsilon = 15\%$, $w = 10$ and various $(\rho_A, \rho_B)$; *left*: $(0.3, 0.7)$, *middle*: $(0.5, 0.5)$, *right*: $(0.7, 0.3)$.

on its proximity to the *AP* and not to feel abused by relaying transit traffic from station *B*.

## VI. CONCLUSIONS

We have proposed a game-theoretic model of a cooperative two-hop wireless relay network where stations can execute SQM attacks to improve received QoS. To discourage such attacks, which easily lead to station *B* starvation, we have designed a reputation-based incentive scheme called RISC2WIN. The scheme instills a noncooperative game between stations *A* and *B*. Provided that both stations attach distinctly more importance to high-priority traffic and assuming plausible threshold-based attack strategies we have demonstrated through simulation that at no $\varepsilon$-Nash equilibrium of the game does starvation occur and moreover, (i) both stations *A* and *B* attack with restraint, (ii) the QoS of intrinsically high-priority traffic depends only on the "natural" competition for the *A*-to-*AP* wireless link between such traffic from stations *A* and *B*, and (iii) station *A* typically does not receive lower QoS for its source traffic than does station *B*. To the best of our knowledge, RISC2WIN is the first cooperative two-hop communication scheme to ensure (i)-(iii) in the presence of SQM attacks. Furthermore, it is easy to implement on top of any wireless technology.

As future work, we plan to analyze other rational strategies (e.g., reinforcement learning, trial-and-error, or regret-based), and convergence to $\varepsilon$-NE under dynamic game scenarios.